\documentclass[a4paper,11pt]{article}
\usepackage[T1]{fontenc} % if needed
\usepackage{float} % if needed

\makeatletter
\gdef\@fpheader{ }
\gdef\@journal{ }
\makeatother
\RequirePackage{amsmath}
\RequirePackage{amssymb}
\RequirePackage{epsfig}
\RequirePackage{graphicx}
\RequirePackage[numbers,sort&compress]{natbib}
\RequirePackage{color}
\RequirePackage[colorlinks=true
,urlcolor=blue
,anchorcolor=blue
,citecolor=blue
,filecolor=blue
,linkcolor=blue
,menucolor=blue
,pagecolor=blue
,linktocpage=true
,pdfproducer=medialab
,pdfa=true
]{hyperref}

\newif\ifnotoc\notocfalse
\newif\ifemailadd\emailaddfalse
\newif\iftoccontinuous\toccontinuousfalse
\makeatletter
\def\@subheader{\@empty}
\def\@keywords{\@empty}
\def\@abstract{\@empty}
\def\@xtum{\@empty}
\def\@dedicated{\@empty}
\def\@arxivnumber{\@empty}
\def\@collaboration{\@empty}
\def\@collaborationImg{\@empty}
\def\@proceeding{\@empty}
\def\@preprint{\@empty}

\newcommand{\subheader}[1]{\gdef\@subheader{#1}}
\newcommand{\keywords}[1]{\if!\@keywords!\gdef\@keywords{#1}\else%
\PackageWarningNoLine{\jname}{Keywords already defined.\MessageBreak Ignoring last definition.}\fi}
\renewcommand{\abstract}[1]{\gdef\@abstract{#1}}
\newcommand{\dedicated}[1]{\gdef\@dedicated{#1}}
\newcommand{\arxivnumber}[1]{\gdef\@arxivnumber{#1}}
\newcommand{\proceeding}[1]{\gdef\@proceeding{#1}}
\newcommand{\xtumfont}[1]{\textsc{#1}}
\newcommand{\correctionref}[3]{\gdef\@xtum{\xtumfont{#1} \href{#2}{#3}}}
\newcommand\jname{JHEP}
\newcommand\acknowledgments{\section*{Acknowledgments}}

\newcommand\preprint[1]{\gdef\@preprint{\hfill #1}}

\makeatother

\newcommand\note[2][]{%
\if!#1!%
\stepcounter{footnote}\footnotetext{#2}%
\else%
{\renewcommand\thefootnote{#1}%
\footnotetext{#2}}%
\fi}

\makeatletter
\newtoks\auth@toks
\renewcommand{\author}[2][]{%
  \if!#1!%
    \auth@toks=\expandafter{\the\auth@toks#2\ }%
  \else
    \auth@toks=\expandafter{\the\auth@toks#2$^{#1}$\ }%
  \fi
}
\makeatother
\makeatletter
\newtoks\affil@toks\newif\ifaffil\affilfalse
\newcommand{\affiliation}[2][]{%
\affiltrue
  \if!#1!%
    \affil@toks=\expandafter{\the\affil@toks{\item[]#2}}%
  \else
    \affil@toks=\expandafter{\the\affil@toks{\item[$^{#1}$]#2}}%
  \fi
}
\makeatother
\makeatletter
\newtoks\email@toks\newcounter{email@counter}%
\newcommand{\emailAdd}[1]{%
\emailaddtrue%
\ifnum\theemail@counter>0\email@toks=\expandafter{\the\email@toks, \@email{#1}}%
\else\email@toks=\expandafter{\the\email@toks\@email{#1}}%
\fi\stepcounter{email@counter}}
\newcommand{\@email}[1]{\href{mailto:#1}{\tt #1}}
\makeatother

\makeatletter
\newcommand*\collaboration[1]{\gdef\@collaboration{#1}}
\newcommand*\collaborationImg[2][]{\gdef\@collaborationImg{#2}}
\makeatletter
\newcommand\afterLogoSpace{\smallskip}
\newcommand\afterSubheaderSpace{\vskip3pt plus 2pt minus 1pt}
\newcommand\afterProceedingsSpace{\vskip21pt plus0.4fil minus15pt}
\newcommand\afterTitleSpace{\vskip23pt plus0.06fil minus13pt}
\newcommand\afterRuleSpace{\vskip23pt plus0.06fil minus13pt}
\newcommand\afterCollaborationSpace{\vskip3pt plus 2pt minus 1pt}
\newcommand\afterCollaborationImgSpace{\vskip3pt plus 2pt minus 1pt}
\newcommand\afterAuthorSpace{\vskip5pt plus4pt minus4pt}
\newcommand\afterAffiliationSpace{\vskip3pt plus3pt}
\newcommand\afterEmailSpace{\vskip16pt plus9pt minus10pt\filbreak}
\newcommand\afterXtumSpace{\par\bigskip}
\newcommand\afterAbstractSpace{\vskip16pt plus9pt minus13pt}
\newcommand\afterKeywordsSpace{\vskip16pt plus9pt minus13pt}
\newcommand\afterArxivSpace{\vskip3pt plus0.01fil minus10pt}
\newcommand\afterDedicatedSpace{\vskip0pt plus0.01fil}
\newcommand\afterTocSpace{\bigskip\medskip}
\newcommand\afterTocRuleSpace{\bigskip\bigskip}
\newlength{\affiliationsSep}
\newcommand\beforetochook{\pagestyle{myplain}\pagenumbering{roman}}

\DeclareFixedFont\trfont{OT1}{phv}{b}{sc}{11}

\renewcommand\maketitle{
\pagestyle{empty}
\thispagestyle{titlepage}
\setcounter{page}{0}
\noindent{\small\scshape\@fpheader}\@preprint\par

\afterLogoSpace
\if!\@subheader!\else\noindent{\trfont{\@subheader}}\fi
\afterSubheaderSpace
\if!\@proceeding!\else\noindent{\sc\@proceeding}\fi
\afterProceedingsSpace
{\LARGE\flushleft\sffamily\bfseries\@title\par}
\afterTitleSpace
\hrule height 1.5\p@%
\afterRuleSpace
\if!\@collaboration!\else
{\Large\bfseries\sffamily\raggedright\@collaboration}\par
\afterCollaborationSpace
\fi
\if!\@collaborationImg!\else
{\normalsize\bfseries\sffamily\raggedright\@collaborationImg}\par
\afterCollaborationImgSpace
\fi
{\bfseries\raggedright\sffamily\the\auth@toks\par}
\afterAuthorSpace
\ifaffil\begin{list}{}{%
\setlength{\leftmargin}{0.28cm}%
\setlength{\labelsep}{0pt}%
\setlength{\itemsep}{\affiliationsSep}%
\setlength{\topsep}{-\parskip}}
\itshape\small%
\the\affil@toks
\end{list}\fi
\afterAffiliationSpace
\ifemailadd %% if emailadd is true
\noindent\hspace{0.28cm}\begin{minipage}[l]{.9\textwidth}
\begin{flushleft}
\textit{E-mail:} \the\email@toks
\end{flushleft}
\end{minipage}
\else %% if emailaddfalse do nothing
\PackageWarningNoLine{\jname}{E-mails are missing.\MessageBreak Plese use \protect\emailAdd\space macro to provide e-mails.}
\fi
\afterEmailSpace
\if!\@xtum!\else\noindent{\@xtum}\afterXtumSpace\fi
\if!\@abstract!\else\noindent{\renewcommand\baselinestretch{.9}\textsc{Abstract:}}\ \@abstract\afterAbstractSpace\fi
\if!\@keywords!\else\noindent{\textsc{Keywords:}} \@keywords\afterKeywordsSpace\fi
\if!\@arxivnumber!\else\noindent{\textsc{ArXiv ePrint:}} \href{http://arxiv.org/abs/\@arxivnumber}{\@arxivnumber}\afterArxivSpace\fi
\if!\@dedicated!\else\vbox{\small\it\raggedleft\@dedicated}\afterDedicatedSpace\fi
\ifnotoc\else
\iftoccontinuous\else\newpage\fi
\beforetochook\hrule
\tableofcontents
\afterTocSpace
\hrule
\afterTocRuleSpace
\fi
\setcounter{footnote}{0}
\pagestyle{myplain}\pagenumbering{arabic}
} % close the \renewcommand\maketitle{

\renewcommand{\baselinestretch}{1.1}\normalsize
\divide\@tempdima\baselineskip
\@tempcnta=\@tempdima
\skip\@mpfootins = \skip\footins
\renewcommand{\@dotsep}{10000}

\newcommand\ps@myplain{
\pagenumbering{arabic}
\renewcommand\@oddfoot{\hfill-- \thepage\ --\hfill}
\renewcommand\@oddhead{}}
\let\ps@plain=\ps@myplain

\newcommand\ps@titlepage{\renewcommand\@oddfoot{}\renewcommand\@oddhead{}}

\numberwithin{equation}{section}

\renewcommand\section{\@startsection{section}{1}{\z@}%
                                   {-3.5ex \@plus -1.3ex \@minus -.7ex}%
                                   {2.3ex \@plus.4ex \@minus .4ex}%
                                   {\normalfont\large\bfseries}}
\renewcommand\subsection{\@startsection{subsection}{2}{\z@}%
                                   {-2.3ex\@plus -1ex \@minus -.5ex}%
                                   {1.2ex \@plus .3ex \@minus .3ex}%
                                   {\normalfont\normalsize\bfseries}}
\renewcommand\subsubsection{\@startsection{subsubsection}{3}{\z@}%
                                   {-2.3ex\@plus -1ex \@minus -.5ex}%
                                   {1ex \@plus .2ex \@minus .2ex}%
                                   {\normalfont\normalsize\bfseries}}
\renewcommand\paragraph{\@startsection{paragraph}{4}{\z@}%
                                   {1.75ex \@plus1ex \@minus.2ex}%
                                   {-1em}%
                                   {\normalfont\normalsize\bfseries}}
\renewcommand\subparagraph{\@startsection{subparagraph}{5}{\parindent}%
                                   {1.75ex \@plus1ex \@minus .2ex}%
                                   {-1em}%
                                   {\normalfont\normalsize\bfseries}}

\def\fnum@figure{\textbf{\figurename\nobreakspace\thefigure}}
\def\fnum@table{\textbf{\tablename\nobreakspace\thetable}}

\long\def\@makecaption#1#2{%
  \vskip\abovecaptionskip
  \sbox\@tempboxa{\small #1. #2}%
  \ifdim \wd\@tempboxa >\hsize
    \small #1. #2\par
  \else
    \global \@minipagefalse
    \hb@xt@\hsize{\hfil\box\@tempboxa\hfil}%
  \fi
  \vskip\belowcaptionskip}

\renewenvironment{thebibliography}[1]{%
\begin{oldthebibliography}{#1}%
\small%
\raggedright%
\setlength{\itemsep}{5pt plus 0.2ex minus 0.05ex}%
}%
{%
\end{oldthebibliography}%
}

\begin{document}

%%%%%%%%%%%%%%%%%%炎籾匈%%%%%%%%%%%%%%%%%%%%%%%%%%%%%&&&&&&&&&&&&&&&&&&&&&&

\title{\boldmath The Brownian Motion in an Ideal Quantum Qas}

% more complex case: 4 authors, 3 institutions, 2 footnotes

\author[a,b,1]{Chi-Chun Zhou,}\note{zhouchichun@dali.edu.cn}
\author[c,b,2]{Ping Zhang,}\note{zhangping@cueb.edu.cn.}
\author[b,3]{and Wu-Sheng Dai}\note{daiwusheng@tju.edu.cn.}

% The "\note" macro will give a warning: "Ignoring empty anchor..."
% you can safely ignore it.

\affiliation[a]{School of Engineering, Dali University, Dali, Yunnan 671003, PR China}
\affiliation[b]{Department of Physics, Tianjin University, Tianjin 300350, PR China}
\affiliation[c]{School of Finance, Capital University of Economics and Business, Beijing 100070, P. R. China}

%\affiliation[c]{DP School}

% e-mail addresses: one for each author, in the same order as the authors
%\emailAdd{Ccc@one.edu.cn}
%\emailAdd{second@asas.edu}
%\emailAdd{daiwusheng@tju.edu.cn}
%\emailAdd{fourth@one.univ}

%\title{\boldmath A title with some math: $x=1$}
%% %simple case: 2 authors, same institution
%% \author{A. Uthor}
%% \author{and A. Nother Author}
%% \affiliation{Institution,\\Address, Country}

% more complex case: 4 authors, 3 institutions, 2 footnotes
%\author[a,b,1]{F. Irst,\note{Corresponding author.}}
%\author[c]{S. Econd,}
%\author[a,2]{T. Hird\note{Also at Some University.}}
%\author[a,2]{and Fourth}

% The "\note" macro will give a warning: "Ignoring empty anchor..."
% you can safely ignore it.

%\affiliation[a]{One University,\\some-street, Country}
%\affiliation[b]{Another University,\\different-address, Country}
%\affiliation[c]{A School for Advanced Studies,\\some-location, Country}

% e-mail addresses: one for each author, in the same order as the authors

%\emailAdd{first@one.univ}
%\emailAdd{second@asas.edu}
%\emailAdd{third@one.univ}
%\emailAdd{fourth@one.univ}

%\date{date}

\abstract{A Brownian particle in an ideal quantum gas is considered. The mean square
displacement (MSD) is derived. The Bose-Einstein or Fermi-Dirac distribution,
other than the Maxwell-Boltzmann distribution, provides a different stochastic
force compared with the classical Brownian motion. The MSD, which depends on
the thermal wavelength and the density of medium particles, reflects the
quantum effect on the Brownian particle explicitly. The result shows that the
MSD in an ideal Bose gas is shorter than that in a Fermi gas. The behavior of
the quantum Brownian particle recovers the classical Brownian particle as the
temperature raises. At low temperatures, the quantum effect becomes obvious.
For example, there is a random motion of the Brownian particle due to the
fermionic exchange interaction even the temperature is near the absolute zero.}
%\keywords{}

\maketitle
\flushbottom
%%%%%%%%%%%%%%%%%%炎籾匈潤崩%%%%%%%%%%%%%%%%%%%%%%%%%%%%%&&&&&&&&&&&&&&&&&&&

%%%%%%%%%%屎猟蝕兵

\section{Introduction}

The Brownian motion is first observed by Robert Brown in 1827\textbf{ }and
then explained by Einstein (1905), Smoluchowski (1905), and Langevin (1908) in
the early 20th century \textbf{\cite{mazo2002brownian}}. The early theory of
the Brownian motion not only provides an evidence for the atomistic hypothesis
of matter \cite{hanggi2005introduction}, but also builds a bridge between the
microscopic dynamics and the macroscopic observable phenomena
\cite{hanggi2005introduction}.

The classical understanding of the Brownian motion is quite well established.
However, there is an assumption in the early theory of the Brownian motion
that the medium particle obeys the Maxwell-Boltzmann distribution.

The behavior of a Brownian particle in an ideal quantum gas draws some
attentions because to study the motion of a Brownian particle in an quantum
system is now within reach of experimental tests. For example, an electron in
a black body radiation \cite{diosi1995quantum}. In such systems, the quantum
exchange interaction, which always leads to real difficulty in mechanics and
statistical mechanics
\cite{dai2017explicit,dai2007interacting,dai2005hard,zhou2018canonical},
exists and causes the medium particle obeying the Bose-Einstein or Fermi-Dirac
distribution. It is difficult to make exact or even detailed dynamical
calculations \cite{bian2016111,mazo2002brownian}, since a different stochastic
force is provided by the Bose-Einstein or Fermi-Dirac distribution. At
high-temperature and low-density, the classical theory serves as good approximation.

\textbf{In this paper,} we give an explicit expression of the mean square
displacement (MSD) of a Brownian particle in an ideal quantum gas using, e.g.,
the virial expansion. Comparison with the classical Brownian motion, a
correction for the MSD, which depends on the thermal wavelength and the
density of medium particles, is deduced. The result shows that the MSD in an
ideal Bose gas is shorter than that in a Fermi gas. The behavior of the quantum
Brownian particle recovers the classical Brownian particle as the temperature
raises. At low temperature, the quantum effect becomes obvious. For example,
there is a random motion of the Brownian particle due to the fermionic
exchange interaction even the temperature is near the absolute zero.

The early studies of the Brownian motion inspired many prominent developments
in various areas such as physics, mathematics, financial markets, and biology.
In physics, exact solutions of Brownian particles in different cases, such as
in a constant field of force \cite{mazo2002brownian} and in a harmonically
potential field \cite{mazo2002brownian}, are given. The Brownian motion
with\ a time dependent diffusion coefficient is studied in Ref.
\cite{bodrova2015ultraslow}. The boundary problem of Brownian motions is
studied in Refs. \cite{huang2015effect,gur2019sensitivity}. The anomalous
diffusion process, frequently described by the scaled Brownian motion, is
studied in Refs. \cite{safdari2015quantifying,jeon2014scaled}. The
Kramers-Klein equation considers the Brownian particle that is in an general
field of force \cite{mazo2002brownian}. The generalized Langevin equations and
the master equation for the quantum Brownian motion are studied in Refs.
\cite{carlesso2017adjoint,weiss2012quantum}. In mathematics, the rigorous
interpretation of Brownian motions based on concepts of random walks,
martingales, and stochastic processes is given
\cite{le2016brownian,bian2016111,mazo2002brownian}. In financial markets, the
theory of the Brownian motion is used to describe the movement of the price of
stocks and options
\cite{bian2016111,mazo2002brownian,kanazawa2018derivation,gairat2017density,kijima2016stochastic,yang2015geometric}%
. The application of the fractional Brownian motion, which is a generalized
Brownian motion, in financial markets is studied in Refs.
\cite{czichowsky2018shadow,rao2016pricing,neuman2018fractional}. Moreover, the
Brownian motion plays a central and fundamental role in studies of soft matter
and biophysics \cite{bian2016111}, e.g., active Brownian motions, which can be
used to describe the motion of swarms of animals in fluid, are studied in
Refs.
\cite{romanczuk2012active,ao2014active,lindner2008diffusion,ebeling2008swarm,pietzonka2016extreme,bian2016111}%
.\

Among many quantities, the MSD, which is measurable, describes the Brownian
motion intuitively.\textbf{ }There are studies focus on the MSD related
problems. For examples, the relation between the MSD and the time interval can
be generally written as $\left\langle x_{t}^{2}\right\rangle \sim t^{\alpha}$
\cite{bodrova2015ultraslow}. One distinguishes the subdiffusion with
$0<\alpha<1$ and the superdiffusion with $\alpha>1$%
\ \cite{bouchaud1990anomalous,metzler2000random,sikora2017mean}. The relation
between the MSD and the time interval of the so called ultraslow Brownian
motions is $\left\langle x_{t}^{2}\right\rangle \sim\left(  \ln t\right)
^{\gamma}$ \cite{metzler2014anomalous}.

There are different approaches to build a quantum analog of the Brownian motion
\cite{lampo2019quantum,grabert1988quantum,caldeira1983path,ambegaokar1991quantum}%
. For examples, the method of the path integral is used to study the quantum
Brownian motion \cite{caldeira1983path}. The approach of a quantum analog or
quantum generalization of the Langevin equation and the master equation, e.g.,
the quantum master equation \cite{diosi1995quantum} and the quantum Langevin
equation \cite{ford1987quantum} is used to build a quantum Brownian motion.
Among them, the method of quantum dynamical semigroups
\cite{lindblad1976generators} is prominent. They point it out that the quantum
equation should be casted into the Lindblad form
\cite{lindblad1976generators,vacchini2000completely}. A completely positive
master equation describing quantum dissipation for a Brownian particle is
derived in Ref. \cite{vacchini2000completely}.

This paper is organized as follows. In Sec. 2, for the sake of completeness,
we derive the brownian motion from the perspective of the particle
distribution in an ideal Boltzmann gas. In Sec. 3, we derive the MSD of a
Brownian particle in an ideal quantum gas. High-temperature and
low-temperature expansions are given. The $d$-dimensional case is considered.
The conclusion and outlook are given in Sec. 4. Some details of the
calculation is given in the Appendix.

\section{A Brownian particle in an ideal classical gas: the Brownian motion}

In this section, we consider a Brownian particle in an ideal classical gas.
For the sake of completeness, we derive, in detail, the Brownian motion from
the perspective of the particle distribution.

\textit{A brief review on the Langevin equation.} For a Brownian particle with
mass $M$, the dynamic equation is given by Paul Langevin
\cite{reichl2009modern}%
\begin{align}
dv  &  =-\frac{\gamma}{M}vdt+\frac{1}{M}F_{t}dt,\label{1}\\
dx  &  =vdt, \label{2}%
\end{align}
where $\gamma=6\pi\eta r$ with $\eta$ the viscous coefficient and $r$ the
radius of the medium particles. $F_{t}$ is the stochastic force generated by
numerous collisions of the medium particle. It is reasonable to make the
assumption that $F_{t}$ is isotropic, i.e.,
\begin{equation}
\left\langle F_{t}\right\rangle =0. \label{16}%
\end{equation}
If the collision of the medium particle is uncorrelated; that is, for $t\neq
s$, $F_{t}$ is independent of $F_{s}$:%
\begin{equation}
\left\langle F_{s}F_{t}\right\rangle \propto\delta\left(  s-t\right)  ,
\label{nnn1}%
\end{equation}
then, the solution of Eqs. (\ref{1}) and (\ref{2}) is \cite{reichl2009modern}%
\begin{align}
v_{t}  &  =v_{0}\exp\left(  -\frac{\gamma}{M}t\right)  +\frac{1}{M}\int%
_{0}^{t}\exp\left[  -\frac{\gamma}{M}\left(  t-s\right)  \right]
F_{s}ds,\label{3}\\
x_{t}  &  =x_{0}+\frac{M}{\gamma}v_{0}\left[  1-\exp\left(  -\frac{\gamma}%
{M}t\right)  \right]  +\frac{1}{\gamma}\int_{0}^{t}\left\{  1-\exp\left[
-\frac{\gamma}{M}\left(  t-s\right)  \right]  \right\}  F_{s}ds, \label{4}%
\end{align}
where $x_{0}$ and $v_{0}$ are the initial position and velocity.

\textbf{The stochastic force determined by the Maxwell-Boltzmann distribution
and the MSD.} In an ideal classical gas, the gas particle obeys the
Maxwell-Boltzmann distribution \cite{pathria2011statistical}. The number of
particles possessing energy within $\varepsilon$ to $\varepsilon+d\varepsilon
$, denoted by $\tilde{a}_{\varepsilon}$, is proportional to $e^{-\beta
\varepsilon}$ \cite{pathria2011statistical}, i.e.,%
\begin{equation}
\tilde{a}_{\varepsilon}=\omega_{\varepsilon}e^{-\beta\varepsilon}d\varepsilon,
\label{nn2}%
\end{equation}
where $\omega_{\varepsilon}$ is the degeneracy of the energy $\varepsilon$ and
$\beta=\left(  kT\right)  ^{-1}$ with $k$ the Boltzmann constant $T$ the
temperature \cite{pathria2011statistical}. A collision of the medium particle
with energy $\varepsilon$ gives a force of magnitude proportional to
$\sqrt{2m\varepsilon}$, which is the momentum of the particle. We have%
\begin{equation}
F=\rho\sqrt{2m\varepsilon}, \label{nn1}%
\end{equation}
where $\rho$ is a coefficient and $m$ is the mass of the medium particle.
Thus, the probability of the Brownian particle subjected to a stochastic force
with magnitude within $F$ to $F+dF$ is%

\begin{equation}
P\left(  F\right)  dF=\sqrt{\frac{\beta}{2\pi m\rho^{2}}}\exp\left(
-\frac{F^{2}\beta}{2m\rho^{2}}\right)  dF. \label{5}%
\end{equation}
In an ideal classical gas, there is no inter-particle interactions among
medium particles, thus, for $t\neq s$, the force $F_{s}$ and $F_{t}$ are
independent. Substituting Eq. (\ref{5}) into Eq. (\ref{nnn1}) gives%
\begin{equation}
\left\langle F_{s}F_{t}\right\rangle =\frac{m\rho^{2}}{\beta}\delta\left(
s-t\right)  . \label{6}%
\end{equation}

By using Eqs (\ref{4}), (\ref{5}), and (\ref{6}), a direct calculation of the
MSD gives%

\begin{align}
\left\langle \left(  x_{t}-x_{0}\right)  ^{2}\right\rangle  &  =\frac{M^{2}%
}{\gamma^{2}}\left(  v_{0}^{2}-\frac{1}{2m\gamma}\frac{m\rho^{2}}{\beta
}\right)  \left[  1-\exp\left(  -\frac{\gamma}{M}t\right)  \right]
^{2}\nonumber\\
&  +\frac{1}{\gamma^{2}}\frac{m\rho^{2}}{\beta}\left\{  t-\frac{M}{\gamma
}\left[  1-\exp\left(  -\frac{\gamma}{M}t\right)  \right]  \right\}  .
\label{7}%
\end{align}
For a large-scale time, $t\gg1$, Eq. (\ref{7}) recovers
\begin{equation}
\left\langle x_{t}^{2}\right\rangle =\frac{kT}{3\pi\eta r}t, \label{8}%
\end{equation}
where $\rho=\sqrt{12\pi\eta r/m}$ and $x_{0}$ is chosen to be $0$ without lose
of generality. Eq. (\ref{8}) is the famous Einstein's long-time result of the
MSD. The motion of a brownian particle in an ideal classical gas is the
Brownian motion.

\section{The MSD of a Brownian particle in an ideal quantum gas}

In this section, we give the MSD of a Brownian particle in an ideal quantum
gas. High-temperature and low-temperature expansions explain the quantum
effect intuitively.

\subsection{The stochastic force determined by the Bose-Einstein or Fermi-Dirac
distribution}

In an\textbf{ }ideal quantum gas, the gas particle obeys Bose-Einstein or
Fermi-Dirac distribution other than the Maxwell-Boltzmann distribution. The
stochastic force is different from that in an ideal classical gas. In this
section, we discuss the properties of the stochastic force in an ideal
quantum gas.

In an\textbf{ }ideal quantum gas, the number of particles possessing energy
within $\varepsilon$ to $\varepsilon+d\varepsilon$, denoted by $a_{\varepsilon
}$, is%

\begin{equation}
a_{\varepsilon}=\frac{\omega_{\varepsilon}}{\exp\left(  \beta\varepsilon
+\alpha\right)  +g}d\varepsilon, \label{14}%
\end{equation}
where $\alpha$ is defined by $z=e^{-\alpha}$ with $z$ the fugacity
\cite{pathria2011statistical}. For Bose cases, $g=-1$, and for Fermi cases,
$g=1$. Thus, the probability of the Brownian particle subjected to a
stochastic force with magnitude within $F$ to $F+dF$ is
\begin{equation}
p\left(  F\right)  dF=\sqrt{\frac{\beta}{2\rho^{2}m\pi}}\frac{1}%
{h_{1/2}\left(  z\right)  }\frac{1}{\exp\left[  \beta F^{2}/\left(  2\rho
^{2}m\right)  +\alpha\right]  +g}dF, \label{15}%
\end{equation}
where we $h_{\nu}\left(  x\right)  $ equals Bose-Einstein integral $g_{\nu
}\left(  x\right)  $ in Bose cases \cite{pathria2011statistical} and
Fermi-Dirac integral $f_{\nu}\left(  x\right)  $ \cite{pathria2011statistical}
in Fermi cases.

In an\textbf{ }ideal quantum gas, the stochastic force is also isotropic, that
is, Eq. (\ref{16}) holds. However, the collision, due to the overlapping of
the wave package, can be correlated; that is, $\left\langle F_{s}%
F_{t}\right\rangle $ is no longer a delta function but a function of $s-t$ with
a peak at $s=t$. However, as the ratio of the thermal wavelength and the
average distance between the medium particles decreases, $\left\langle
F_{s}F_{t}\right\rangle $, can be well approximated by a delta function:
\begin{equation}
\left\langle F_{s}F_{t}\right\rangle \sim\frac{m\rho^{2}}{\beta}\frac
{h_{3/2}\left(  z\right)  }{h_{1/2}\left(  z\right)  }\delta\left(
s-t\right)  , \label{17}%
\end{equation}
for $n\lambda\ll1$, where $\lambda=h/\sqrt{2\pi mkT}$ is the thermal
wavelength and $n$ is the density of the medium particle.

In this paper, we consider the case that the ratio of the thermal wavelength
and the average distance between the medium particles is small.

\subsection{The MSD}

For $n\lambda\ll1$, by using Eqs. (\ref{15}), (\ref{16}), and (\ref{4}), a
direct calculation of MSD gives%
\begin{align}
\left\langle x_{t}^{2}\right\rangle  &  =\frac{M^{2}}{\gamma^{2}}\left\{
v_{0}^{2}-\frac{1}{2m\gamma}\frac{m\rho^{2}}{\beta}\frac{h_{3/2}\left(
z\right)  }{h_{1/2}\left(  z\right)  }\right\}  \left[  1-\exp\left(
-\frac{\gamma}{M}t\right)  \right]  ^{2}\nonumber\\
&  +\frac{1}{\gamma^{2}}\frac{m\rho^{2}}{\beta}\frac{h_{3/2}\left(  z\right)
}{h_{1/2}\left(  z\right)  }\left\{  t-\frac{M}{\gamma}\left[  1-\exp\left(
-\frac{\gamma}{M}t\right)  \right]  \right\}  . \label{18}%
\end{align}
For a large-scale time, $t\gg1$, Eq. (\ref{18}) recovers%

\begin{equation}
\left\langle x_{t}^{2}\right\rangle =\frac{kT}{3\pi\eta r}t\frac
{h_{3/2}\left(  z\right)  }{h_{1/2}\left(  z\right)  }, \label{9}%
\end{equation}
where $h_{3/2}\left(  z\right)  /h_{1/2}\left(  z\right)  $ is a correction
for the MSD due to the Bose-Einstein or Fermi-Dirac distribution, a result of
the quantum exchange interaction among gases particles.

\subsection{High-temperature and low-temperature expansions}

In order to compare with Eq. (\ref{8}) intuitively, we give high-temperature
and low-temperature expansions of Eq. (\ref{9}) by using the state equation of
ideal Bose or Fermi gases\textbf{
\cite{reichl2009modern,pathria2011statistical} }%
\begin{align}
p  &  =\frac{kT}{\lambda}h_{3/2}\left(  z\right)  ,\label{n1}\\
\frac{N}{V}  &  =\frac{1}{\lambda}h_{1/2}\left(  z\right)  . \label{n02}%
\end{align}

\textbf{The high-temperature expansion. }At high temperatures, the virial
expansion of Eqs. (\ref{n1}) and (\ref{n02}) directly gives
\cite{reichl2009modern,pathria2011statistical}%
\begin{equation}
\frac{pV}{N}\sim kT\left[  1+ga_{1}\left(  T\right)  n\lambda+...\right]  ,
\label{n03}%
\end{equation}
where $a_{1}\left(  T\right)  $ is the first virial coefficient
\cite{reichl2009modern}. For a $1$-dimensional ideal Bose or Fermi gas,
$a_{1}\left(  T\right)  =0.353553$ \cite{pathria2011statistical}. Substituting
Eqs. (\ref{n1}) and (\ref{n02}) into Eq. (\ref{n03}) gives
\begin{equation}
\frac{h_{3/2}\left(  z\right)  }{h_{1/2}\left(  z\right)  }\sim\left[
1+ga_{1}\left(  T\right)  n\lambda+...\right]  . \label{n04}%
\end{equation}

Substituting Eq. (\ref{n04}) into Eq. (\ref{9}) gives the MSD at high
temperatures:%
\begin{equation}
\left\langle x_{t}^{2}\right\rangle =\frac{kT}{3\pi\eta r}t\left[
1+ga_{1}\left(  T\right)  n\lambda+\ldots\right]  . \label{11}%
\end{equation}
The result, Eq. (\ref{11}), shows that the MSD in an ideal Bose gas is shorter
than that in a Fermi gas. Since $\lambda$ decreases as $T$ raises, the behavior
of the quantum Brownian particle returns the classical Brownian particle as
the temperature raises.

\textbf{The low-temperature expansion for Fermi cases. }At low temperatures,
for Fermi cases, $g=-1$,
\begin{equation}
\frac{h_{3/2}\left(  z\right)  }{h_{1/2}\left(  z\right)  }=\frac
{f_{3/2}\left(  z\right)  }{f_{1/2}\left(  z\right)  }. \label{n05}%
\end{equation}
The expansion of the Fermi-Dirac integral at large $z$ gives
\cite{pathria2011statistical}%
\begin{equation}
f_{\nu}\left(  e^{\xi}\right)  =\frac{\xi^{\nu}}{\Gamma\left(  1+\nu\right)
}\left\{  1+2\nu\sum_{j=1,3,5,...}\left[  \left(  \nu-1\right)  ....\left(
\nu-j\right)  \left(  1-2^{-j}\right)  \frac{\zeta\left(  j+1\right)  }%
{\xi^{j+1}}\right]  \right\}  . \label{n06}%
\end{equation}
Keeping only the first two terms in Eq. (\ref{n06}) gives%
\begin{equation}
f_{\nu}\left(  z\right)  =\frac{\left(  \ln z\right)  ^{\nu}}{\Gamma\left(
1+\nu\right)  }+2\nu\left(  \nu-1\right)  \frac{1}{2}\frac{\zeta\left(
2\right)  }{\left(  \ln z\right)  ^{2}}. \label{n07}%
\end{equation}
Substituting Eq. (\ref{n07}) into Eqs. (\ref{n1}) and (\ref{n02}) gives%
\begin{align}
p  &  =\frac{kT}{\lambda}\frac{\left(  \ln z\right)  ^{3/2}}{\Gamma\left(
5/2\right)  }\left[  1+\frac{3}{4}\frac{\zeta\left(  2\right)  }{\left(  \ln
z\right)  ^{2}}\right]  ,\label{n08}\\
\frac{N}{V}  &  =\frac{1}{\lambda}\frac{\left(  \ln z\right)  ^{1/2}}%
{\Gamma\left(  5/2\right)  }\left[  1-\frac{1}{4}\frac{\zeta\left(  2\right)
}{\left(  \ln z\right)  ^{2}}\right]  . \label{n09}%
\end{align}
The fugacity $z$ can be solved from Eq. (\ref{n09}):
\begin{equation}
\ln z\sim\frac{\epsilon_{f}}{kT}\left[  1+\frac{1}{2}\zeta\left(  2\right)
\left(  \frac{kT}{\epsilon_{f}}\right)  ^{2}\right]  , \label{n10}%
\end{equation}
where $\epsilon_{f}=\lambda^{2}kT\left[  \frac{1}{2}\Gamma\left(  \frac{3}%
{2}\right)  n\right]  ^{2}$ is the Fermi energy \cite{pathria2011statistical}.
By substituting Eq. (\ref{n07}) into Eq. (\ref{n05}) with fugacity $z$ given by
Eq. (\ref{n10}), we have%
\begin{equation}
\frac{f_{3/2}\left(  z\right)  }{f_{1/2}\left(  z\right)  }=\frac
{\Gamma\left(  3/2\right)  }{\Gamma\left(  5/2\right)  }\frac{\epsilon_{f}%
}{kT}\left\{  1+\left[  \frac{\zeta\left(  2\right)  }{2}+\zeta\left(
2\right)  \right]  \left(  \frac{kT}{\epsilon_{f}}\right)  ^{2}\right\}  .
\label{n11}%
\end{equation}

Substituting Eq. (\ref{n11}) into Eq. (\ref{9}) gives the MSD of Fermi cases
at low temperatures:%
\begin{equation}
\left\langle x_{t}^{2}\right\rangle \sim\frac{1}{3\pi\eta r}t\frac
{\Gamma\left(  3/2\right)  }{\Gamma\left(  5/2\right)  }\epsilon_{f}\left[
1+\frac{3}{2}\zeta\left(  2\right)  \left(  \frac{kT}{\epsilon_{f}}\right)
^{2}\right]  . \label{12}%
\end{equation}
The first term of Eq. (\ref{12}) is independent of the temperature $T$, which
means that there is a random motion of the Brownian particle due to the
fermionic exchange interaction even the temperature is near the absolute zero.
It is a result of Pauli exclusion principle \cite{pathria2011statistical}.

\textbf{The low-temperature expansion for Bose cases. }At low temperatures,
for Bose cases, $g=1$,%

\begin{equation}
\frac{h_{1+d/2}\left(  z\right)  }{h_{d/2}\left(  z\right)  }=\frac
{g_{1+d/2}\left(  z\right)  }{g_{d/2}\left(  z\right)  }. \label{n13}%
\end{equation}
Expanding $g_{\nu}\left(  z\right)  $ around $z=1$ gives
\cite{pathria2011statistical}%
\begin{equation}
g_{\nu}\left(  z\right)  =\frac{\Gamma\left(  1-\nu\right)  }{\left(  -\ln
z\right)  ^{1-\nu}}+\sum_{j=0}^{\infty}\frac{\left(  -1\right)  ^{j}}{j!}%
\zeta\left(  \nu-j\right)  \left(  -\ln z\right)  ^{j}. \label{n14}%
\end{equation}
Substituting Eq. (\ref{n14}) into Eqs. (\ref{n1}) and (\ref{n02}) gives%

\begin{align}
p  &  =\frac{1}{\lambda^{d}}\Gamma\left(  -\frac{1}{2}\right)  \left(  -\ln
z\right)  ^{1/2}+\zeta\left(  \frac{3}{2}\right)  -\zeta\left(  \frac{1}%
{2}\right)  \left(  -\ln z\right)  ,\label{n15}\\
\frac{N}{V}  &  =\frac{1}{\lambda^{d}}\frac{\Gamma\left(  1/2\right)
}{\left(  -\ln z\right)  ^{1/2}}+\zeta\left(  \frac{1}{2}\right)
-\zeta\left(  -\frac{1}{2}\right)  \left(  -\ln z\right)  . \label{n16}%
\end{align}
The fugacity can be solved from Eq. (\ref{n16}):
\begin{equation}
\ln z=-\frac{\pi}{n^{2}\lambda^{2}}. \label{n17}%
\end{equation}
By substituting Eq. (\ref{n14}) into Eq. (\ref{n13}) with fugacity $z$ given by
Eq. (\ref{n17}), we have%

\begin{equation}
\frac{g_{3/2}\left(  z\right)  }{g_{1/2}\left(  z\right)  }=\frac{\zeta\left(
3/2\right)  }{\sqrt{n^{2}\lambda^{2}}}-\left(  2+\frac{\zeta\left(
3/2\right)  \zeta\left(  1/2\right)  }{\pi}\right)  \frac{\pi}{n^{2}%
\lambda^{2}}. \label{n18}%
\end{equation}

Substituting Eq. (\ref{n18}) into Eq. (\ref{9}) gives the MSD of Bose cases at
low temperatures:%

\begin{align}
\left\langle x_{t}^{2}\right\rangle  &  \sim\frac{kT}{3\pi\eta r}t\left[
\zeta\left(  \frac{3}{2}\right)  \frac{1}{n\lambda}-\left(  2\pi+\zeta\left(
\frac{3}{2}\right)  \zeta\left(  \frac{1}{2}\right)  \right)  \frac{1}%
{n^{2}\lambda^{2}}\right] \label{13}\\
&  \sim\frac{1}{3\pi\eta r}\zeta\left(  \frac{3}{2}\right)  \frac{\sqrt{2\pi
m}}{h}\frac{1}{n}\left(  kT\right)  ^{3/2}t.
\end{align}
The MSD is proportional to $T^{3/2}$ and is reversely proportional to the
density of particle, which is also different from that of the Brownian motion.

\subsection{The $d$-dimensional case}

In this section, a similar procedure gives the MSD of a Brownian particle in a
$d$-dimensional space. For the sake of clarity, we list the result. The detail
of the calculation can be found in the Appendix.

\textbf{The MSD. }The MSD for a Brownian particle in an ideal quantum gas in a
$d$-dimensional space is%

\begin{equation}
\left\langle \mathbf{x}_{t}^{2}\right\rangle =\frac{kTd}{3\pi\eta r}%
t\frac{h_{1+d/2}\left(  z\right)  }{h_{d/2}\left(  z\right)  }. \label{n77}%
\end{equation}

\textbf{The high-temperature expansion. }At high temperatures, the MSD,
Eq.(\ref{n77}), becomes\textbf{ }%
\begin{equation}
\left\langle \mathbf{x}_{t}^{2}\right\rangle =\frac{kTd}{3\pi\eta r}t\left[
1+ga_{1}\left(  T\right)  n\lambda^{d}+\ldots\right],  \label{20}%
\end{equation}
where $a_{1}\left(  T\right)  =\frac{1}{2^{1+d/2}}$ and is the first virial
coefficient of ideal Bose or Fermi gases in a $d$-dimensional space
\cite{reichl2009modern,pathria2011statistical}.

\textbf{The low-temperature expansion for Fermi cases. }At low temperatures,
for Fermi cases, the MSD, Eq. (\ref{n77}), becomes%
\begin{equation}
\left\langle \mathbf{x}_{t}^{2}\right\rangle \sim\frac{d}{3\pi\eta r}%
\frac{\Gamma\left(  1+d/2\right)  }{\Gamma\left(  2+d/2\right)  }t\epsilon
_{f}\left[  1+\left(  \frac{d}{2}+1\right)  \zeta\left(  2\right)  \left(
\frac{kT}{\epsilon_{f}}\right)  ^{2}\right]  , \label{21}%
\end{equation}
where $\epsilon_{f}$ is the Fermi energy in a $d$-dimensional space and
$\epsilon_{f}=\lambda^{2}kT\left[  \frac{1}{2}\Gamma\left(  1+\frac{d}%
{2}\right)  n\right]  ^{2/d}$ \cite{reichl2009modern,pathria2011statistical}.

\textbf{The low-temperature expansion for Bose cases with }$d=2$\textbf{. }The
low-temperature expansion for a Bose gas at any dimension higher than $2$ is
not given in this section, because the Bose-Einstein condensation (BEC)
occurs. Here, we only consider the $2$-dimensional case.\textbf{ }At low
temperatures, for Bose cases, the MSD, Eq. (\ref{n77}), becomes
\begin{equation}
\left\langle \mathbf{x}_{t}^{2}\right\rangle =\frac{2kT}{3\pi\eta r}t\left[
-\exp\left(  -n\lambda^{2}\right)  -\frac{\exp\left(  -n\lambda^{2}\right)
}{n\lambda^{2}}+\frac{\pi^{2}}{6n\lambda^{2}}\right].  \label{22}%
\end{equation}

\section{Conclusions and outlooks}

The difficulty in the calculation of the behavior of a Brownian particle in an
ideal quantum gas directly comes from the stochastic force caused by the
Bose-Einstein and Fermi-Dirac distribution other than the Maxwell-Boltzmann
distribution. Comparison with the classical Brownian motion, on one hand, the
distribution of the stochastic force is different; on the other hand, the
collision, due to the overlapping of the wave package, could be correlated,
that is, $\left\langle F_{s}F_{t}\right\rangle $ is no longer a delta function
but a function of $s-t$ with a peak at $s=t$. Thus, it is difficult to make
exact or even detailed dynamical calculations
\cite{bian2016111,mazo2002brownian}.

In this paper, we consider the motion of a Brownian particle in an ideal
quantum gas. We give an explicit expression of the MSD, which depends on the
thermal wavelength and the density of medium particles. High-temperature and
low-temperature expansions explain the quantum effect intuitively. For
examples, the MSD in an ideal Bose gas is shorter than that in a Ferm gas.
There is a random motion of the Brownian particle due to the fermionic
exchange interaction even the temperature is near the absolute zero.

The result in this work can be verified by experiment test.

\section{Acknowledgments}
We are very indebted to Dr G Zeitrauman for his encouragement. This work is supported
in part by NSF of China under Grant No. 11575125 and No. 11675119.

%\appendix
%\section{Some title}
%Please always give a title also for appendices.

\section{Appendix}

In the appendix, we give the detail of the calculation of Eqs (\ref{n77}),
(\ref{20}), (\ref{21}), and (\ref{22}).

\textbf{The detail of the calculate for the MSD, Eq. (\ref{n77}), of a
Brownian particle in a }$d$\textbf{-dimensional space.} The Langevin equation
in $d$-dimensional is%
\begin{align}
M\frac{d\mathbf{v}}{dt}  &  =-\gamma\mathbf{v}+\mathbf{F}_{t},\label{n2}\\
\frac{d\mathbf{x}}{dt}  &  =\mathbf{v.} \label{n3}%
\end{align}
In a $d$-dimensional space, the stochastic force $\mathbf{F}_{t}$ is
isotropic:%
\begin{equation}
\left\langle \mathbf{F}\right\rangle =0. \label{n4}%
\end{equation}
For different time $t$ and $s$, $\mathbf{F}_{s}$ and $\mathbf{F}_{t}$ are
almost independent when the ratio of the thermal wavelength and the average
distance between the medium particles is small, that is,%
\begin{equation}
\left\langle \mathbf{F}_{s}\cdot\mathbf{F}_{t}\right\rangle \sim\delta\left(
s-t\right)  \label{nn5}%
\end{equation}
holds for $n\lambda^{d}\ll1$. The solution of Eqs. (\ref{n2}) and (\ref{n3})
is
\begin{align}
\mathbf{v}_{t}  &  =\mathbf{v}_{0}e^{-\frac{\gamma}{M}t}+\frac{1}{M}\int%
_{0}^{t}\exp\left[  -\frac{\gamma}{M}\left(  t-s\right)  \right]
\mathbf{F}_{s}ds,\label{26}\\
\mathbf{x}_{t}  &  =\mathbf{x}_{0}+\frac{M}{\gamma}v_{0}\left[  1-\exp\left(
-\frac{\gamma}{M}t\right)  \right]  +\frac{1}{\gamma}\int_{0}^{t}\left\{
1-\exp\left[  -\frac{\gamma}{M}\left(  t-s\right)  \right]  \right\}
\mathbf{F}_{s}ds. \label{27}%
\end{align}

In the $d$-dimensional case, the number of particle possessing momentum within
$\mathbf{P}$ to $\mathbf{P+}d\mathbf{P}$, denoted by $a\left(  \mathbf{P}%
\right)  $, is \cite{reichl2009modern,pathria2011statistical}
\begin{equation}
a\left(  \mathbf{P}\right)  =\frac{1}{\exp\left[  \beta\left(  p_{x^{1}}%
^{2}+p_{x^{2}}^{2}+...p_{x^{d}}^{2}\right)  /\left(  2m\right)  +\alpha
\right]  +g}. \label{23}%
\end{equation}
\qquad The force given by a collision of a particle with momentum $\mathbf{P}$
is proportional to $\mathbf{P}$, $\mathbf{F=}\rho\mathbf{P}$. Thus the
probability of the stochastic force with magnitude within $\left\vert
\mathbf{F}\right\vert $ to $\left\vert \mathbf{F+}d\mathbf{F}\right\vert $ is%
\begin{equation}
P\left(  \mathbf{F}\right)  d\mathbf{F}=\left(  \sqrt{\frac{\beta}{2\pi
m\rho^{2}}}\right)  ^{d}\frac{1}{h_{d/2}\left(  z\right)  }\frac{1}%
{\exp\left[  \beta\left\vert \mathbf{F}\right\vert ^{2}/\left(  2m\rho
^{2}\right)  +\alpha\right]  +g}. \label{24}%
\end{equation}
Substituting Eq. (\ref{24}) into Eq. (\ref{nn5}) gives
\begin{equation}
\left\langle \mathbf{F}_{s}\cdot\mathbf{F}_{t}\right\rangle \sim\frac
{dm\rho^{2}}{\beta}\frac{h_{1+d/2}\left(  z\right)  }{h_{d/2}\left(  z\right)
}\delta\left(  s-t\right)  . \label{25}%
\end{equation}

By using Eqs. (\ref{27}), (\ref{24}), and (\ref{25}), a direct calculation of
MSD gives%
\begin{align}
\left\langle \mathbf{x}_{t}^{2}\right\rangle  &  =\frac{M^{2}}{\gamma^{2}%
}\left[  \mathbf{v}_{0}^{2}-\frac{d}{2m\gamma}\frac{m\rho^{2}}{\beta}%
\frac{h_{1+d/2}\left(  z\right)  }{h_{d/2}\left(  z\right)  }\right]  \left[
1-\exp\left(  -\frac{\gamma}{M}t\right)  \right]  ^{2}\nonumber\\
&  +\frac{1}{\gamma^{2}}\frac{dm\rho^{2}}{\beta}\frac{h_{1+d/2}\left(
z\right)  }{h_{d/2}\left(  z\right)  }\left[  t-\frac{M}{\gamma}\left[
1-\exp\left(  -\frac{\gamma}{M}t\right)  \right]  \right]  . \label{n6}%
\end{align}
where $\mathbf{x}_{0}$ is chosen to be the origin.

For a large-scale time, $t\gg1$, Eq. (\ref{n6}) recovers Eq. (\ref{n77}).

\textbf{The high-temperature expansion}. For the $d$-dimensional case, the
state equation of an ideal quantum gas is
\cite{reichl2009modern,pathria2011statistical}%
\begin{align}
p  &  =\frac{kT}{\lambda^{d}}h_{1+d/2}\left(  z\right)  ,\label{A001}\\
\frac{N}{V}  &  =\frac{1}{\lambda^{d}}h_{d/2}\left(  z\right)  . \label{A002}%
\end{align}
The virial expansion \cite{reichl2009modern,pathria2011statistical} directly
gives%
\begin{equation}
\frac{pV}{N}\sim kT\left[  1+ga_{1}\left(  T\right)  n\lambda^{d}+...\right]
\label{A01}%
\end{equation}
Substituting Eqs. (\ref{A001}) and (\ref{A002}) into Eq. (\ref{A01}) gives
\begin{equation}
\frac{h_{1+d/2}\left(  z\right)  }{h_{d/2}\left(  z\right)  }\sim\left[
1+ga_{1}\left(  T\right)  n\lambda^{d}+...\right]  . \label{A03}%
\end{equation}
Substituting Eq. (\ref{A03}) into Eq. (\ref{n77}) gives Eq. (\ref{20}).

\textbf{The low-temperature expansion for Fermi cases}. For Fermi cases,
$g=-1$,
\begin{equation}
\frac{h_{1+d/2}\left(  z\right)  }{h_{d/2}\left(  z\right)  }=\frac
{f_{1+d/2}\left(  z\right)  }{f_{d/2}\left(  z\right)  }. \label{A0}%
\end{equation}
By the expansion of the Fermi-Dirac integral at large $z$, we have
\cite{pathria2011statistical}%
\begin{equation}
f_{\nu}\left(  e^{\xi}\right)  =\frac{\xi^{\nu}}{\Gamma\left(  1+\nu\right)
}\left\{  1+2\nu\sum_{j=1,3,5,...}\left[  \left(  \nu-1\right)  ....\left(
\nu-j\right)  \left(  1-2^{-j}\right)  \frac{\zeta\left(  j+1\right)  }%
{\xi^{j+1}}\right]  \right\}  .
\end{equation}
Keeping only the first two terms gives%
\begin{equation}
f_{\nu}\left(  z\right)  =\frac{\left(  \ln z\right)  ^{\nu}}{\Gamma\left(
1+\nu\right)  }+2\nu\left(  \nu-1\right)  \frac{1}{2}\frac{\zeta\left(
2\right)  }{\left(  \ln z\right)  ^{2}}. \label{A3}%
\end{equation}
Substituting Eq. (\ref{A3}) into Eqs. (\ref{A001}) and (\ref{A002}) gives%
\begin{align}
\frac{p}{kT}  &  =\frac{1}{\lambda^{d}}\frac{\left(  \ln z\right)  ^{1+d/2}%
}{\Gamma\left(  2+d/2\right)  }\left[  1+\frac{d}{2}\left(  1+\frac{d}%
{2}\right)  \frac{\zeta\left(  2\right)  }{\left(  \ln z\right)  ^{2}}\right]
,\\
N  &  =\frac{\Omega}{\lambda^{d}}\frac{\left(  \ln z\right)  ^{d/2}}%
{\Gamma\left(  1+d/2\right)  }\left[  1+\frac{d}{2}\left(  \frac{d}%
{2}-1\right)  \frac{\zeta\left(  2\right)  }{\left(  \ln z\right)  ^{2}%
}\right]  , \label{A1}%
\end{align}
where $\Omega$ is the volume. The fugacity can be solved from Eq (\ref{A1}):
\begin{equation}
\ln z\sim\frac{\epsilon_{f}}{kT}\left[  1-\zeta\left(  2\right)  \left(
\frac{d}{2}-1\right)  \left(  \frac{kT}{\epsilon_{f}}\right)  ^{2}\right]  ,
\label{A2}%
\end{equation}
where $\epsilon_{f}=\lambda^{2}kT\left[  \frac{1}{2}\Gamma\left(  1+\frac
{d}{2}\right)  n\right]  ^{2/d}$ is the Fermi energy. By substituting Eq.
(\ref{A3}) into Eq. (\ref{A0}) with fugacity $z$ given by Eq. (\ref{A2}), we
have%
\begin{equation}
\frac{f_{1+d/2}\left(  z\right)  }{f_{d/2}\left(  z\right)  }=\frac
{\Gamma\left(  1+\frac{d}{2}\right)  }{\Gamma\left(  2+\frac{d}{2}\right)
}\frac{\epsilon_{f}}{kT}\left\{  1+\left[  \frac{d\zeta\left(  2\right)  }%
{2}+\zeta\left(  2\right)  \right]  \left(  \frac{kT}{\epsilon_{f}}\right)
^{2}\right\}  . \label{A4}%
\end{equation}
Substituting Eq. (\ref{A4}) into Eq. (\ref{n77}) gives Eq. (\ref{21}).

\textbf{The low-temperature expansion for Bose cases in the }$2$%
\textbf{-dimensional space.} For Bose cases, $g=1$,%

\begin{equation}
\frac{h_{2}\left(  z\right)  }{h_{1}\left(  z\right)  }=\frac{g_{2}\left(
z\right)  }{g_{1}\left(  z\right)  },\label{A11}%
\end{equation}
where $d=2$. For $d=2$,
\begin{equation}
g_{1}\left(  z\right)  =-\ln\left(  1-z\right)  .\label{A08}%
\end{equation}
Substituting Eq. (\ref{A08}) into Eq. (\ref{A002}) gives%
\begin{equation}
\frac{N}{V}=-\frac{1}{\lambda^{2}}\ln\left(  1-z\right)  .\label{A07}%
\end{equation}
Then, the fugacity can be solved from Eq (\ref{A07}):%

\begin{equation}
z=1-\exp\left(  -n\lambda^{2}\right)  .\label{A13}%
\end{equation}
Expanding $g_{2}\left(  z\right)  $ around $z=1$ gives%
\begin{align}
g_{2}\left(  z\right)   &  =\frac{\pi^{2}}{6}-\left(  1-z\right)
-\frac{\left(  1-z\right)  ^{2}}{2^{2}}-\frac{\left(  1-z\right)  ^{3}}{3^{2}%
}-\ldots\nonumber\\
&  +\left(  1-z\right)  \ln\left(  1-z\right)  +\frac{\left(  1-z\right)
^{2}}{2}\ln\left(  1-z\right)  +\frac{\left(  1-z\right)  ^{3}}{3}\ln\left(
1-z\right)  +\ldots\label{A144}%
\end{align}
Substituting Eqs. (\ref{A144}) and (\ref{A08}) with fugacity given in Eq.
(\ref{A13}) into Eq. (\ref{A11}) gives%
\begin{equation}
\frac{g_{2}\left(  z\right)  }{g_{1}\left(  z\right)  }=-\exp\left(
-n\lambda^{2}\right)  -\frac{\exp\left(  -n\lambda^{2}\right)  }{n\lambda^{2}%
}+\frac{\pi^{2}}{6n\lambda^{2}},\label{A14}%
\end{equation}
Substituting Eq. (\ref{A14}) into Eq. (\ref{n77}) gives Eq. (\ref{22}).

\acknowledgments

We are very indebted to Dr G. Zeitrauman for his encouragement. This work is supported in part by NSF of China under Grant No.
11575125 and No.  11675119.

%\acknowledgments%崑仍
%%%%%%%%%%屎猟潤崩

%\begin{thebibliography}{99}

%\end{thebibliography}\endgroup

%\bibitem{a}
%Author, \emph{Title}, \emph{J. Abbrev.} {\bf vol} (year) pg.

%\bibitem{b}
%Author, \emph{Title},
%arxiv:1234.5678.

%\bibitem{c}
%Author, \emph{Title},
%Publisher (year).

% Please avoid comments such as "For a review'', "For some examples",
% "and references therein" or move them in the text. In general,
% please leave only references in the bibliography and move all
% accessory text in footnotes.

% Also, please have only one work for each \bibitem.

%\end{thebibliography}

\bibliographystyle{JHEP}
\bibliography{refs}% Produces the bibliography via BibTeX.

\end{document}